\documentclass[doublecol,figures]{epl2} %for 2 columns style without line numbers

\usepackage[T1]{fontenc}
\usepackage[utf8]{inputenc}
\usepackage{amsmath,amsfonts,amssymb,bbm}

\usepackage{ulem}
\usepackage{cancel}

\newcommand\be{\begin{equation}}
\newcommand\ee{\end{equation}}

\title{Probabilistic picture for particle number densities in stretched tips of the branching Brownian motion}
\shorttitle{Probabilistic picture for particle number densities}

\author{A. D. Le \inst{1}\thanks{Email: \email{dung.le@polytechnique.edu}}
    \and A. H. Mueller\inst{2}\thanks{Email: \email{amh@phys.columbia.edu}}
    \and S. Munier\inst{1}\thanks{Email: \email{stephane.munier@polytechnique.edu}}
    }
\shortauthor{A. D. Le, A. H. Mueller, S. Munier}
\institute{
  \inst{1}
  CPHT, CNRS, \'Ecole polytechnique, IP Paris,
  F-91128 Palaiseau, France\\
  \inst{2}Department of Physics, Columbia University,
  New York, NY 10027, USA
}


\abstract{
  In the framework of a stochastic picture for the one-dimensional branching Brownian motion, we compute the probability density of the number of particles near the rightmost one at a time $T$, that we take very large, when this extreme particle is conditioned to arrive at a predefined position $x_T$ chosen far ahead of its expected position $m_T$. We recover the previously-conjectured fact that the typical number density of particles a distance $\Delta$ to the left of the lead particle, when both $\Delta$ and $x_T-\Delta-m_T$ are large, is smaller than the mean number density by a factor proportional to $e^{-\zeta\Delta^{2/3}}$, where $\zeta$ is a constant that was so far undetermined. Our picture leads to an expression for the probability density of the particle number, from which a value for $\zeta$ may be inferred.
}

\begin{document}

\maketitle

\section{Introduction}

The branching Brownian motion (BBM) \cite{IkedaNagasawaWatanabe.1968} is one of the simplest representatives of a class of stochastic processes that model phenomena appearing in a large diversity of contexts \cite{Brunet:2016}. Evolutionary biology may be the most natural one \cite{Murray:2002}, but branching models have also proved relevant in less expected fields, such as e.g. economics \cite{Benhabib:2020}. It is the subject of very active studies in mathematics \cite{Bovier:2017}.

Our initial interest in the BBM actually came from {high-energy} physics. The quantum states of {elementary} particles interacting at very high energies, which determine the scattering cross sections measured at colliders, can be thought of as being generated {by} a peculiar branching process \cite{Mueller:1993rr} that may be mapped to the BBM in the appropriate limit \cite{Munier:2003vc,Mueller:2014fba}. Universal properties derived for the BBM may thus be taken over to observables relevant for {collider} physics (see Refs.~\cite{Munier:2009pc,Le:2021hel} for reviews).

{In this paper, we will call ``particle'' any object completely characterized by a single real number $x$ (its spatial position) and which evolves in time by the BBM: its position obeys a Brownian motion, and it may be replaced, at an exponentially distributed time, by two particles at the same position, which subsequently evolve by independent BBMs. We start with one particle at $x=0$ at the initial time $t=0$.} For definiteness, we set the diffusion constant of the Brownian motion to $\frac12$ and the branching rate to unity, in such a way that the expectation value $\bar n(T,x)$ of the density $N(T,x)$ of these particles at position $x$ at a given evolution time $T$ reads
\be
\bar n(T,x)\equiv \left\langle N(T,x)\right\rangle=\frac{1}{\sqrt{2\pi T}}e^{T-{x^2}/{(2T)}}.
\ee
This completely defines the process.

The extremal regions of the BBM, close to the lead particles at time~$T$, turn out to be of particular interest. The probability density of the position $R_T$ of, say, the rightmost particle in a realization is deduced from the probability $u(t,x)$ that there be at least one particle to the right of the position $x$ at time $t$, a quantity that obeys the Fisher-Kolmogorov-Petrovsky-Piscunov (FKPP) equation~\cite{Fisher:1937,KPP:1937} in the form
\be
\partial_t u=\frac12\partial^2_x u+u-u^2\,,
\label{eq:FKPP}
\ee
with the initial condition $u(0,x)={\mathbbm{1}}_{\{x\leq 0\}}$. The probability to find $R_T$ in the interval $[x,x+dx]$ then obviously reads
\be
   {\mathbbm P}(R_T\in dx) = \tilde u(T,x) dx,
   \ \text{with}\
   \tilde u(T,x) \equiv -\partial_x u(T,x).
\ee
The leading terms in the large-$T$ expansion of the expectation value of $R_T$ are known rigorously~\cite{Bramson:1983}:
\be
m_T\equiv\langle R_T\rangle=\sqrt{2}\,T-\frac{3}{2\sqrt{2}}\ln T+ {c_m+o(1)},
\ee
{where $c_m$ is an undetermined constant.} An analytic form is also known for $\tilde u$ {(see e.g. Ref.~\cite{BrunetDerrida:1997,Ebert:2000})},
\be
\tilde u(T,x)\simeq c(x-m_T)e^{-\sqrt{2}(x-m_T)}\exp\left({-\frac{(x-m_T)^2}{2T}}\right),
\label{eq:tildeu}
\ee
that is valid for $T\gg 1$ and\footnote{{A less strict upper bound on $x-m_T$ was established in Ref.~\cite{DerridaMeersonSasorov:2016}, but the more conservative one~(\ref{eq:scaling-region}) will be enough for our purpose.}}
\be
1\ll x-m_T{\leq {\cal O}(\sqrt{T})}\,.
\label{eq:scaling-region}
\ee
$c$ in eq.~(\ref{eq:tildeu}) is another undetermined constant. Note that when the rightmost inequality in eq.~(\ref{eq:scaling-region}) is replaced by the {strong ordering} $x-m_T\ll\sqrt{T}$, the Gaussian factor in eq.~(\ref{eq:tildeu}) drops out, and $\tilde u(T,x)$ then only depends on the single variable $x-m_T$: therefore, we call this range in $x$ the ``scaling region''.

{Let us denote by $N(\Delta t,\Delta x)$ the random density of particles generated by the BBM evolved over the time interval $\Delta t$ starting with one single particle, evaluated at the relative position $\Delta x$ with respect to that of the initial particle.}
As shown in Ref.~\cite{BrunetDerrida:2009,BrunetDerrida:2011}, all statistical properties of {$N$} near the lead particle may be deduced from solutions to the FKPP equation (\ref{eq:FKPP}) with appropriate initial conditions. {We will focus on ``typical values'' of these particle densities, that we shall quantify with the help of the median of their probability distributions, and compare to first moments.}

In Ref.~\cite{Mueller:2019ror}, the {density of particles} at a large time $T$ at a fixed distance $\Delta$ behind the lead particle was studied {in two cases: for the unconditioned BBM, and when the lead particle at final time is required to have a given position.} In the case of the unconditioned BBM, the mean and typical {densities of particles turn out to be of the same order for large $\Delta$, namely
\be
\begin{aligned}
    {\left\langle N(T,R_T-\Delta)\right\rangle}&\propto{n_\text{typical}(T,R_T-\Delta)},\\
  \text{and}\ \ n_\text{typical}(T,R_T-\Delta) &\propto\Delta\, e^{\sqrt{2}\Delta}.
  \end{aligned}
\label{eq:ntypicalevol_res}
\ee
We shall attribute to the notation ``$\propto$'' the meaning of ``asymptotic proportionality'': the ratio of the left and right-hand sides of this binary operator tends to some finite constant when the relevant variable ($\Delta$ here) becomes large.} When, instead, the rightmost particle is conditioned to arrive at the predefined position $R_T=x_T$ (see fig.~\ref{fig:illustration} for an illustration) chosen deep in the scaling region,  {namely when the tip is conditioned to extend far to the right,} then the mean {densities of particles} and the typical ones were found to scale {widely differently from one another}:
\be
\begin{split}
{\langle N(T,R_T-\Delta|R_T=x_T)\rangle}& {\propto} e^{\sqrt{2}\Delta},\\
n_\text{typical}(T,R_T-\Delta|R_T=x_T)& {\propto} e^{\sqrt{2}\Delta -\zeta\Delta^{2/3}}.
\end{split}
\label{eq:n_res}
\ee
The exact value of the constant of order unity $\zeta$ appearing in the expression of $n_\text{typical}$ could not be determined in the calculation {of Ref.~\cite{Mueller:2019ror}}. {These expressions sign a very skewed probability density towards low numbers of particles, that contribute very little to the first moments.}

The formulas~(\ref{eq:ntypicalevol_res}),(\ref{eq:n_res}) should hold for asymptotically-large values of $\Delta$, and provided that $x_T-\Delta$ is also in the scaling region; this, in turn, requires the evolution time $T$ to be much larger than $\Delta^2$.

Our interest in this particular observable stemmed from the so-called ``phenomenological model for front fluctuations'', proposed for stochastic FKPP fronts which appear in the study of the BBM with selection~\cite{Brunet:2005bz}, and subsequently applied to the BBM without selection (and to branching random walks) \cite{Mueller:2014gpa}. The central observation was that the main source of stochasticity in the position of the final lead particle\footnote{As for the case of the BBM without selection, the position of the lead particle we are referring to is that measured in the Lalley and Sellke frame \cite{Lalley:1987}.} consisted in rare but large fluctuations that would intermittently send a few particles significantly ahead of the expected position of the lead particle, and that these few particles could actually be treated as a single one, meaning that they can be characterized by a single position variable. The resulting effective particle was produced at some random intermediate time, and once produced, evolved further deterministically. Thus, in the phenomenological model, it was assumed without proof that the particle number density a distance $\Delta$ behind the lead particle in such fluctuations was much smaller than the density at this same distance $\Delta$ from the lead in a typical evolution. The results~(\ref{eq:ntypicalevol_res}),(\ref{eq:n_res}) confirmed that this hypothesis was correct, and quantified the diluteness of the particles in these tip fluctuations.

The analytical formulas~(\ref{eq:n_res}) followed from an intricate heuristic analysis of the solution to the FKPP equation with peculiar initial conditions \cite{Mueller:2019ror} {(at variance with eqs.~(\ref{eq:ntypicalevol_res}), which are much more straightforward to establish)}. The purpose of this Letter is to rederive the typical density {of particles} in a stochastic picture, which turns out to be much more intuitive. In addition, this picture leads to a determination of the value of $\zeta$, and even of the probability of the density {of particles}.

Our starting point consists in the exact description proposed in Ref.~\cite{Brunet:2020yiq} of the BBM conditioned in such a way that its lead particle at final time is found at some predefined position. We then turn to a simplified model of this formulation, {for the sake of being able} to obtain quantitative {asymptotic} results.

%%%%%%%%%%%%%%%%%%%%%%%%%%%%%%%%%%%

\section{Exact description of the conditioned BBM}

\begin{figure}
\onefigure[width=\columnwidth]{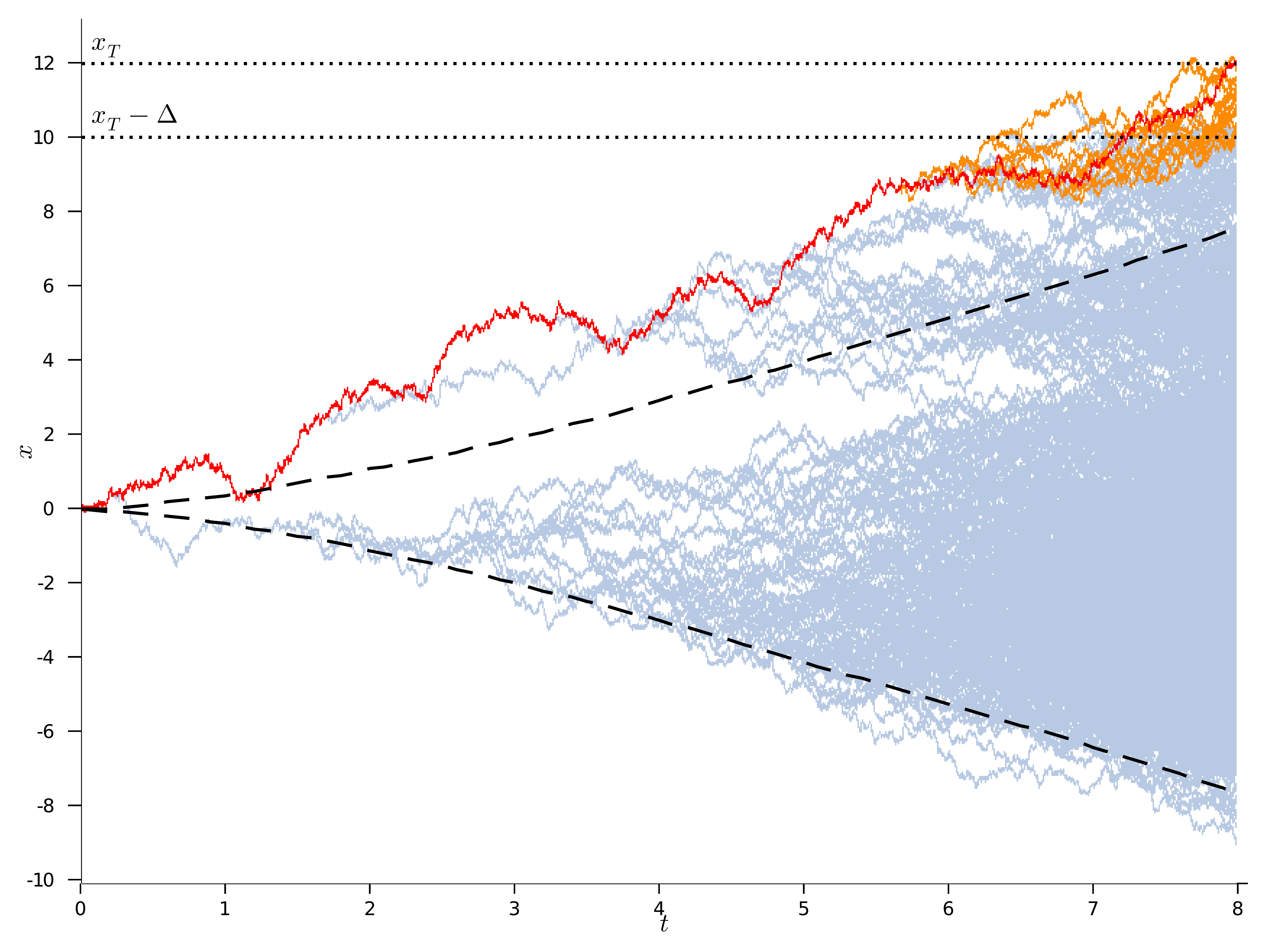}
\caption{An exact realization of the BBM conditioned to have its lead particle at position $x_T=12$ at the final time $T=8$. The trajectory of the latter is shown in red. The trajectories of all particles emitted off the red one that have their rightmost offspring in the interval $[x_T-\Delta,x_T)$ at time $T$ are displayed in orange ($\Delta=2$ in this figure). The expectation values of the trajectories of the rightmost and leftmost lead particles, of equations $x=\pm m_t$, are represented in dashed lines.}
\label{fig:illustration}
\end{figure}

Let us consider {the BBM conditioned in such a way that the lead particle at time $T$ has position $R_T=x_T$}. This particle has a unique ancestor at any given time $t\leq T$: we call its random position $R_t$. We single out the realizations of the spacetime trajectories $(t,x=R_t)$ by coloring them in red (see fig.~\ref{fig:illustration}). Hence by definition, in a given realization of the {thus conditioned} BBM, the red particle is the one whose rightmost offspring at the final time $T$ is at position {$x_T$}.

Let us pick a particle at time $t$ and assume that it is found at the position $x$. The probability that the position of its rightmost offspring at the final time be not less than $x_T$ reads $U(t,x)\equiv u(T-t,x_T-x)$, and the probability that its rightmost offspring end up in the infinitesimal interval $[x_T,x_T+dx_T]$ reads $\tilde U(t,x)dx_T$, with
\be
\tilde U(t,x)\equiv\tilde u(T-t,x_T-x).
\label{eq:tildeU}
\ee

It turns out that when the lead particle is required to arrive at the predefined position $x_T$ at final time, which is the case on which we shall focus in this Letter, the realizations of the trajectories $(t,x=R_t)$ of the red particle are that of a Brownian motion with the spacetime-dependent drift $\partial_x\ln\tilde U(t,x)$. The reader is refered to~\cite{Brunet:2020yiq} for a derivation.\footnote{Note that the idea of trading a constraint on a process for an effective force has also been proposed as a method to generate realizations of (non-branching) Brownian motions subject to global dynamical constraints; see Ref.~\cite{De_Bruyne:2021}.}

The effect of this peculiar drift on a Brownian particle can easily be understood qualitatively, from the functional form of $\tilde U$. One may easily check that as one moves out of a region of size $\sqrt{T-t}$ around the position $x_T-m_{T-t}$ in which $\ln \tilde U$ is finite, $\ln\tilde U$ becomes negative and eventually goes to $-\infty$. It actually drops fast enough for its $x$-derivative (namely the drift) to become large and positive (resp. negative) on the left (resp. on the right) of that region. Hence, when due to the Brownian motion, the red particle deviates from $x_T-m_{T-t}$ by typically more than $\sqrt{T-t}$, it experiences a restoring force that drives it back. When $t$ approaches the final time $T$, this region becomes narrower, in such a way that the red particle be appropriately directed to its final position $x_T$.

Let us introduce the probability density $p(t,x)$ of the position of the red particle at time $t$: ${\mathbbm{P}}(R_t\in dx)\equiv p(t,x)dx$. From general considerations (see e.g. \cite{Gardiner:2004}), $p$ obeys the Fokker-Planck equation
\be
\partial_t p=\frac12\partial_x^2 p-\partial_x(\partial_x \ln\tilde U\,p).
\label{eq:FP}
\ee

The ``non-red'' particles, that we define as those which have their rightmost offspring at time $T$ at positions strictly less than $x_T$, stem from branchings of the red particle into ``red + non-red'' pairs. The probability for such a branching to occur in the time interval $[t,t+dt]$, when the red particle is at position {$x$}, reads \cite{Brunet:2020yiq}
\begin{multline}
  {\mathbbm{P}}(\text{red}\rightarrow\text{red+non-red}\in dt|R_t={x})\\
  =2[1-U(t,{x})]dt.
   \label{eq:branching}
\end{multline}
The non-red particles evolve {by} a branching Brownian motion. {The defining conditioning to not overtake the position $x_T$ at final time is implemented by a specific space-time-dependence of the processes: the Brownian motion} has unit variance and spacetime-dependent drift $\partial_x\ln [1-U(t,x)]$, and the {branchings occur at the rate $1-U(t,x)$} \cite{Brunet:2020yiq}.

The set of rules we have just stated is a formulation of the spine description {\cite{Hardy:2009}} of the conditioned BBM. Let us show how we may take advantage of this formulation to compute the typical number of particles in a stretched tip.

%%%%%%%%%%%%%%%%%%%%%%%%%%%%%%%%%%%

\section{{Asymptotics of the density of particles at the final time $T$}}

{The random density of particles at the position $x_T-\Delta$ in the conditioned BBM can be decomposed as
  \begin{multline}
    N(T,R_T-\Delta|R_T=x_T)=\\
    \sum_{i=1}^{M(T)} N^{(i)}(T-t^{(i)},x_T-\Delta-R_{t^{(i)}}|R_{T-t^{(i)}}^{(i)}<x_T-R_{t^{(i)}}),
    \label{eq:decomposition}
  \end{multline}
where the sum goes over the random number $M(T)$ of branchings of the red particle on the random trajectory $\{R_t\}$ into red+non-red pairs, occurring at the random times $\{t^{(1)},\cdots,t^{(i)},\cdots,t^{(M(T))}\}$. The $N^{(i)}$'s stand for the densities of particles resulting from the respective stochastic evolutions of these non-red particles over the time interval $T-t^{(i)}$ from their creations, at relative position $x_T-\Delta-R_{t^{(i)}}$. For each of these evolutions, $R^{(i)}_{T-t^{(i)}}$ is the final position of the particle that ends up in the lead in its respective tree, relative to the position $R_{t^{(i)}}$ of the root of the latter.}

{In the following, we shall} focus on very large times {$T\gg 1$, and} pick the positions $x_T$ and $x_T-\Delta$ in the scaling region, with the further condition $\Delta\gg 1$. We conjecture that {the following simplifying assumptions can be made without altering} the behavior of {the typical density of particles in the conditioned BBM} in these asymptotic limits: {{\it (i)} the fluctuations of the densities of particles $N^{(i)}$ can be neglected, and the sum over $i$ in eq.~(\ref{eq:decomposition}) may be replaced by an integration over time weighted by the intensity of the ``red$\rightarrow$red+non-red'' point process (``mean-field'' approximation); {\it (ii)} one may ignore the conditioning in $N^{(i)}$.}
{Under these assumptions,} the probability density of $N$ is in direct correspondence with that of the trajectory of the red particle: {$N$ conditioned to a realization $\{r_t\}$ of the process $\{R_t\}$, with $r_T=x_T$, is a deterministic function of time and position which reads
  \begin{multline}
    N(T,R_T-\Delta|\{R_{t<T}=r_t,R_T=x_T\})\simeq\\
    \int{\mathbbm{P}}(\text{red}\rightarrow\text{red+non-red}\in dt|R_t=r_t)\\
    \times n_\text{typical}(T-t,x_T-\Delta-r_t).
    \label{eq:N0}
  \end{multline}
}
We shall first discuss the statistical properties of the trajectory {of the red particle, from which we will be able to justify further crucial approximations. We will then turn to the evaluation of $n_\text{typical}$ under the assumptions $(i)$ and $(ii)$, from which we will be able to arrive at a closed expression for $N$ conditioned as in eq.~(\ref{eq:N0}); see eq.~(\ref{eq:n_final}) below. We will finally deduce the probability density of $N$ from the statistics of the trajectory of the red particle, with the help of the explicit expression of the density of particles for one particular trajectory of the red particle thus obtained; see eq.~(\ref{eq:distribution}) below.}

%%%

\subsection{Statistics of the trajectory of the red particle}

Let us exhibit an analytical approximation of the solution to the Fokker-Planck equation~(\ref{eq:FP}). We know the expression of the drift term around the scaling region. Indeed, inserting eq.~(\ref{eq:tildeu}) into eq.~(\ref{eq:tildeU}) and taking the derivative of its logarithm, we get
\be
\partial_x \ln\tilde U(t,x) {\simeq} -\frac{1}{x_t-x}+\sqrt{2}+\frac{x_t-x}{T-t},
\label{eq:partial_x_ln_tilde_U}
\ee
where we have introduced the notation
\be
x_t\equiv x_T-m_{T-t}.
\ee

Anticipating that the relevant values of $x$ will be such that $x_t-x={\cal O}(\sqrt{T-t})$, we may keep in $x_t$ only the two largest terms when $T-t$ is large, namely replace it by
\be
x_t^0\equiv x_T-\sqrt{2}(T-t),
\label{eq:def_xt0}
\ee
{neglecting a term proportional to $\ln(T-t)$.}
Within this approximation, the Fokker-Planck equation maps to that of a three-dimensional Bessel bridge (see e.g. \cite{Revuz:2013}). We check that the {probability} density $p=p^0$, where
\be
p^{0}(t,x)\equiv \sqrt{\frac{2}{\pi}}
\frac{\left(x_t^{0}-x\right)^2}{(T-t)^{3/2}}\exp\left({-\frac{\left(x_t^{0}-x\right)^2}{2(T-t)}}\right),
\label{eq:p0}
\ee
is an exact solution to eq.~(\ref{eq:FP}) when $\partial_x\ln \tilde U$ is replaced by eq.~(\ref{eq:partial_x_ln_tilde_U}) in which we substitute $x_t^0$ for $x_t$.

When $t$ approaches the final time $T$, the probability density $p^0$ becomes concentrated at $x_T$. Starting with a single particle at $t=0$, we expect eq.~(\ref{eq:p0}) to be valid for all trajectories with significant probability and for times ordered as $T-t\ll t$.

It will prove useful to parametrize the trajectory of the red particle in terms of the rescaled random variable
\be
\Xi_t\equiv\frac{x_t-R_t}{\sqrt{T-t}}.
\ee
The replacement $x_t\rightarrow x_t^0$ defines a new random variable $\Xi_t^0$, the law of which is read off eq.~(\ref{eq:p0}):
\be
   {\mathbbm{P}}(\Xi_t^0\in d\xi^0)={\sqrt{\frac{2}{\pi}}}\,\left(\xi^{0}\right)^2 e^{-\left(\xi^0\right)^2/2}d\xi^0\,.
   \label{eq:density_Xi}
\ee
This law is manifestly independent of time, and the typical values taken by $\Xi_t^0$ are of order unity at all times.
%%%

We see that {the position} $R_t$ {of the red particle} does not come very close to $x_t^0$ with sizable probability, thanks to the quadratic prefactor in its density~(\ref{eq:p0}). This means that we may assume that $U(t,R_t)\ll 1$ for all relevant values of $t$. Hence, in eq.~(\ref{eq:N0}), we can consider that the red particle branches out non-red particles as a Poisson point process in time of constant intensity~2, {leading to the simple formula
  \begin{multline}
    N(T,R_T-\Delta|\{R_{t<T}=r_t,R_T=x_T\})\simeq\\
    2 \int_0^T dt\, n_\text{typical}(T-t,x_T-\Delta-r_t).
    \label{eq:N}
  \end{multline}
}

%%%

\subsection{{Typical density of non-red particles from a single evolution}}
The function $n_\text{typical}$ appearing in  {eqs.~(\ref{eq:N0}),}~(\ref{eq:N}) is the {typical density of particles generated by an unconstrained BBM, evaluated at a finite time.} {Following an original idea of Brunet and Derrida (see Ref. \cite{BrunetDerrida:1997}) for the BBM with selection, applied to the BBM without selection in Ref.~\cite{Mueller:2014gpa},\footnote{{See eq.~(6) in Ref.~\cite{Mueller:2014gpa}, and Sec.~4.1 therein for a heuristic calculation leading to such an expression.}} $n_\text{typical}$ can be estimated by solving a partial differential equation of the type
\be
\partial_t n_\text{typical}=\frac12\partial_x^2 n_\text{typical}+n_\text{typical}\times\Theta(n_\text{typical})\,,
\ee
starting from a localized initial condition around $x=0$, with $\int dx\, n_\text{typical}(0,x)=1$. The ``cutoff function'' $\Theta$ is such that $\Theta(n)\to 1$ for $n\gg 1$ and $\Theta(n)$ goes rapidly to zero when $n$ decreases below~1. This cutoff is meant to implement the fact there is no particle to the right of some position in realizations,} and was shown to account correctly for the leading effect of discreteness of the particle numbers, i.e. of stochasticity.\footnote{See Ref.~\cite{MuellerMytnikQuastel:2011} for a mathematical formulation {in the case of the BBM with selection}.} {(The detailed form of $\Theta$ is unimportant; $\Theta(n)={\mathbbm{1}}_{\{n\geq 1\}}$ is a possible choice \cite{BrunetDerrida:1997}, as well as $\Theta(n)=n/(1+n)$ \cite{Mueller:2014gpa}.)} When the initial particle is at position $x$ at time $t<T$, the typical {density of particles} at position $y$ at time $T$ {is then seen to read}
\begin{multline}
n_\text{typical}(T-t,y-x){\simeq} c'(x+m_{T-t}-y)e^{-\sqrt{2}[y-(x+m_{T-t})]}\\
\times\exp\left(-\frac{[y-(x+m_{T-t})]^2}{2(T-t)}\right)
{{\mathbbm{1}}_{\left\{x+m_{T-t}-y\geq 0\right\}}}.
\label{eq:n_typical}
\end{multline}
{The indicator function vanishes at the position of the Brunet-Derrida cutoff, which logically, turns out to coincide with the expectation value of the position of the rightmost particle.} $c'$ is an undetermined constant of order unity. {Note that one recovers the expression in eq.~(\ref{eq:ntypicalevol_res}) when the evolution time is sent to infinity, limit in which the Gaussian factor becomes unity. However, for the present calculation, it is crucial to keep these finite-time corrections: indeed, while $T$ is sent to infinity, the typical time elapsed from the creation of the non-red particle from its branching off the red one, $T-t$, which coincides with the variance of the Gaussian in eq.~(\ref{eq:n_typical}), will turn out to be large, but finite.}

%%%

\subsection{{Evaluation of $N$ for a given trajectory of the red particle}}
{The next step consists in inserting~(\ref{eq:n_typical}) into~(\ref{eq:N}), and performing the integration over time.}
We identify $y$ {in eq.~(\ref{eq:n_typical})} to the position at which we measure the number of particles, namely $x_T-\Delta$. Then, $y-m_{T-t}=x_t-\Delta$. Consistently with the approximation made for the trajectory of the red particle, {which amounted to neglecting $\ln(T-t)$ compared to $T-t$,} we may replace $x_t$ by $x_t^0$ {(namely $y-m_{T-t}$ by $x_t^0-\Delta$)} in all factors in eq.~(\ref{eq:n_typical}), except in the first exponential: {the latter enhances indeed that logarithm to a power. Therefore,} we set $y-m_{T-t}=x_t^0-\Delta+\frac{3}{2\sqrt{2}}\ln (T-t)$ {therein.} We also put the initial particle on a realization of the trajectory of the red particle by identifying $x$ to a realization of $R_t$. Parametrizing the latter with the help of the rescaled random variable $\Xi_t$, which we replace by $\Xi_t^0$ for the consistency of the approximations, and considering one particular realization $\xi_t^0$ of it, we find that the typical number density reads
\begin{multline}
  n_\text{typical}(T-t,y-x){\simeq}c'\frac{\Delta-\xi_t^0\sqrt{T-t}}{(T-t)^{3/2}}
  e^{\sqrt{2}(\Delta-\xi_t^0\sqrt{T-t})}\\
  \times\exp\left[{-\frac12\left(\xi_t^0-\frac{\Delta}{\sqrt{T-t}}\right)^2}\right]
  {{\mathbbm{1}}_{\left\{T-t\leq \left({\Delta}/{\xi_{t}^0}\right)^2\right\}}}.
\end{multline}
Inserting this expression into eq.~(\ref{eq:N}), we find, for such a particular realization,
\begin{multline}
  {N}(T,x_T-\Delta|\{\xi^0_t\}){\simeq}\\
  2c'\int_0^{{T}} \left(\Delta-\xi_{{T-\tilde t}}^0\sqrt{\tilde t}\right) e^{\sqrt{2}\left(\Delta-\xi_{{T-\tilde t}}^0\sqrt{\tilde t}\right)}\\
  \times\exp\left[{-\frac12\left(\xi_{{T-\tilde t}}^0-\frac{\Delta}{\sqrt{\tilde t}}\right)^2}\right]
  {\mathbbm{1}}_{\left\{\tilde{t}\leq {\left(\Delta/\xi_{{T-\tilde t}}^0\right)^2}\right\}}\,
  \frac{d\tilde t}{\tilde t^{3/2}},
\label{eq:Nint}
\end{multline}
where $\tilde t\equiv T-t$.  {(The conditioning on a given trajectory of the red particle ending up at $x_T$ has been expressed in terms of the rescaled trajectory, and the notation has been simplified in a self-explanatory way).}

%%%

We finally need to evaluate the integral that appears in eq.~(\ref{eq:Nint}). On a given trajectory of the red particle, in the limit of large $\Delta$, the integral is dominated by a saddle point, located at the time $\tilde t_s$ that extremizes the sum of the arguments of the exponentials in the integrand of eq.~(\ref{eq:Nint}). {We assume that $\xi_t^0$ is constant} in a large-enough region around the location of the saddle point, and small compared to $\Delta/\sqrt{\tilde t_s}$. Then, the saddle-point equation is solved by
\be
\tilde t_s=T-t_s=\frac{\Delta^{4/3}}{2^{1/3}(\xi_{t_s}^0)^{2/3}}.
\ee
{The constancy of $\xi_t^0$ over the relevant integration region is instrumental: we will check this assumption {\it a posteriori}}. We see that $\Delta/\sqrt{\tilde t_s}=2^{1/6}(\xi_{t_s}^0\Delta)^{1/3}$. Keeping in mind that $\xi^0_{t_s}={\cal O}(1)$, it indeed dominates over $\xi_{t_s}^0$: this justifies that we could neglect the latter compared to the former. Expanding the integrand {in eq.~(\ref{eq:Nint})} around $\tilde t_s$, we get
\begin{multline}
  {N}(T,x_T-\Delta|\{\xi_t^0\})
  {\simeq}
  2c'\frac{\Delta}{\tilde t_s^{3/2}} e^{\sqrt{2}\Delta-\frac{3}{2^{2/3}}(\xi_{t_s}^0\Delta)^{2/3}}\\
    \times \int_0^{{\min}\left[{T},\left(\Delta/\xi^0_{t_s}\right)^2\right]} \exp\left[{-\frac34\left(\frac{\xi^0_{t_s}}{\Delta}\right)^2(\tilde t-\tilde t_s)^2}\right]d\tilde t.
\label{eq:Nsp}
\end{multline}

We are now in a position to check the crucial {\it a priori} hypothesis of the constancy of $\xi_t^0$. {Since the integral~(\ref{eq:Nsp}) is that of a Gaussian of variance of order $\Delta^2$ centered at $\tilde t_s$,} the integration region is effectively a region of size $\Delta$ around $\tilde t_s$. {Since the red particle obeys a Bessel process, its position varies by typically $\sqrt{\Delta}$ in that region. Hence its rescaled position} $\xi_t^0$ varies only by the negligible amount ${{\cal O}\left(\sqrt{\Delta/(T-t_s)}\right)}\sim 1/\Delta^{1/6}$: this justifies that we may indeed keep $\xi_t^0$ fixed to $\xi^0_{t_s}$ {in the integral~(\ref{eq:Nsp})}.
{We can also argue that assumption {\it (ii)} above is correct. The non-red particles created at the times that effectively contribute to the integral in~(\ref{eq:Nsp}) (i.e. around $t_s$, time at which their rescaled position is $\xi^0_{t_s}$) have their typical lead particle at final time a distance ${\cal O}(\sqrt{\tilde t_s})\sim\Delta^{2/3}\gg 1$ to the left of $x_T$. Thus, the requirement that the lead of the non-red particles ends up to the left of $x_T$ turns out to be satisfied without the imposition of any conditioning. We conclude that non-red particles essentially evolve by the unconditioned BBM, as assumed in {\it (ii)}.}

Finally, we perform the integral over $\tilde t$ in eq.~(\ref{eq:Nsp}). The lower and upper integration bounds may be set to $-\infty$ and $+\infty$ respectively without altering {neither} the {large-$T$ nor the} large-$\Delta$ asymptotics. We are then led to the following expression:
\be
{N}(T,x_T-\Delta|\{\xi_t^0\})
{\simeq}
4\sqrt\frac{2\pi}{3}c'\, e^{\sqrt{2}\Delta-\left[\frac{3}{2^{2/3}}(\xi_{t_s}^0)^{2/3}\right]\Delta^{2/3}}.
\label{eq:n_final}
\ee
This is the density of particles yielded by the particular trajectory parametrized by $\xi_{t}^0$. The probability density of the random variable $\Xi^0_t$ of which $\xi_t^0$ is a realization is given by eq.~(\ref{eq:density_Xi}) and is seen to have a width of order unity, meaning that the coefficient of $\Delta^{2/3}$ varies significantly from event-to-event.

%%%

\subsection{Probability of the {density of particles}}

From the density {of particles conditioned on} a given trajectory of the red particle (\ref{eq:n_final}) together with the probability density~(\ref{eq:density_Xi}) of its rescaled position $\Xi_{t_s}^0$ at the saddle point, we can deduce the probability of the number density of particles at position $x_T-\Delta$. Writing $\bar n\equiv e^{\sqrt{2}\Delta}$, which up to an overall constant factor, is the mean {density of particles} for large $\Delta$ appearing in eq.~(\ref{eq:n_res}), 
\be
   {\mathbbm{P}}\left(\frac{\ln \left({{\bar n}/{N}}\right)}{\Delta^{2/3}}\in dl\right)
   {\simeq}
   \frac{4}{27}\sqrt{\frac{2}{3\pi}}\,{l^{{7}/{2}}}\,
   e^{-{2}l^3/27}dl
   \label{eq:distribution}
\ee
in the logarithmic scale that is relevant here, and expressed with the help of the appropriate scaling variable. It is clear in this explicit formula that the number of particles fluctuates widely between realizations.

{The typical density of particles can be estimated by computing the median of the probability law~(\ref{eq:distribution}). We find eq.~(\ref{eq:n_res})  for $n_\text{typical}$, with the constant $\zeta$ solving} the implicit equation
\be
\frac{\Gamma\left(\frac32,\frac{2}{27}\zeta^3\right)}{\Gamma(\frac32)}=\frac12,
\ee
where $\Gamma(s,x)\equiv\int_x^\infty t^{s-1}e^{-t}dt$ is the incomplete gamma function. Solving numerically this equation, we find $\zeta=2.5182\cdots$.

%%%%%%%%%%%%%%%%%

\section{Discussion and outlook}

We have recovered the functional form~(\ref{eq:n_res}) of the typical particle number density near the tip of a BBM whose lead particle at final time is conditioned to have its position deep in the scaling region. While the original derivation of eq.~(\ref{eq:n_res}), exposed in Ref.~\cite{Mueller:2019ror}, relied on the analysis of the solutions to the FKPP equation, the present approach uses a probabilistic picture expected to capture the features of a spine formulation. If the model we use here is accurate enough, as we expect, we have bridged the gap in the original calculation of Ref.~\cite{Mueller:2019ror} by providing a value for the constant $\zeta$ appearing in eq.~(\ref{eq:n_res}). However, owing to the huge spread in the almost-equally probable values of the particle number density encoded in the probability density $p^0$ (eq.~(\ref{eq:p0})), the precise value of this number might not be a very relevant quantity in this problem. But through our calculation, we actually get an expression for a more detailed quantity: the probability of the particle number density. Therefore, our main result is eq.~(\ref{eq:distribution}), or, alternatively, eq.~(\ref{eq:n_final}) together with eq.~(\ref{eq:density_Xi}).

Since several assumptions that we are not in a position to justify rigorously at this point were needed to arrive at the latter expression, it would be desirable to have a numerical check of the statistics of $N$. But given that the relative weights of the terms neglected through the different approximations leading to eq.~(\ref{eq:distribution}) take the form of small negative powers of the parameter $\Delta$, the latter has to be taken extremely large for the subasymptotic terms to vanish, and hence also the evolution time $T$ has to be huge. Therefore, accurate numerical calculations, to which e.g. the probability (\ref{eq:distribution}) may be compared, are not possible yet, even if one uses dedicated algorithms such as the one introduced in Ref.~\cite{Brunet:2020yiq}. We shall leave these investigations for future work.

{On the mathematical side, it is known that the point measure of the particle positions near the tip of the BBM converges, at asymptotic times, to a {decorated Poisson point process} with exponential intensity~\cite{BrunetDerrida:2011,Arguin:2012,Arguin:2013}. By forcing the tip of realizations of the BBM to stretch to large distances to the right of the expected position $m_T$ of the rightmost particle, we may be isolating single realizations of the decoration. (Note that this decoration may be the appropriate mathematical object that represents what was called ``tip fluctuations'' in Ref.~\cite{Mueller:2014gpa}). Our result might then amount to an expression for the typical number of particles generated a distance $\Delta$ from the maximum $x_T$ of realizations of the decoration process. The latter was explicitly described using a backward spine construction~\cite{Aidekon:2013,CortinesHartungLouidor:2019}. This backward construction seems to bear a lot of similarities with the forward construction on which the present work builds (see also Ref.~\cite{Berestycki:2022}). In particular, a three-dimensional Bessel process was found to rule the backward spine. We hope that a precise link can eventually be established, so that one may transpose our discussion into the rigorous backward formulation: this may be the key to a mathematical proof of our heuristic results.
}

\acknowledgments
SM thanks the organizers of the workshop ``Branching systems, reaction-diffusion equations and population models'' held at Centre de Recherches Math\'ematiques, Universit\'e de Montr\'eal (May 2022), for the opportunity to present the research that forms the basis of this report. He also thanks the participants for their interest and for many stimulating discussions.  {We are grateful to the anonymous referees for their very careful critical reading, and for their numerous insightful suggestions, which have helped to improve our work and especially to make it more accessible to the mathematical community.}

The work of ADL and SM was partially supported by the Agence Nationale de la Recherche under the project ANR-16-CE31-0019. The  work  of  AHM  is  supported in part by the U.S. Department of Energy Grant DE-FG02-92ER40699.

%\bibliographystyle{eplbib}
%\bibliography{biblio}

\end{document}